\documentclass[preprint2,twoside]{hwo}

\usepackage{lipsum}

\input{hwo.h}

\usepackage{enumitem}
\begin{document}

\title{\textbf{\LARGE System Analysis for a high-precision high-accuracy Astrometric instrument for HWO}}
\author {\textbf{\large Jérôme Amiaux,$^{1}$ Fabien Malbet,$^{2}$ Florence Ardellier-Desages,$^{1}$ Eric Doumayrou,$^{1}$ Pierre-Antoine Frugier,$^{1}$ Renaud Goullioud,$^{4}$ Thomas Greene,$^{5}$ Lucas Labadie,$^{6}$ Pierre-Olivier Lagage,$^{1}$ Manon Lizzana,$^{3,10}$ Alain Leger,$^{7}$ Thierry Lepine,$^{8}$ Gary Mamon,$^{9}$ Jérôme Martignac,$^{1}$ Julien Michelot,$^{10}$ Fabrice Pancher,$^{2}$ Thibault Pichon,$^{1}$ Aki Roberge,$^{11}$ Samuel Ronayette,$^{1}$ Hugo Rousset,$^{2}$ Breann Sitarski,$^{11}$ Alessandro Sozzetti,$^{12}$ Thierry Tourette$^{1}$}}
\affil{$^1$\small\it Univ. Paris-Saclay, Univ. Paris Cité, CEA, CNRS, AIM, France}
\affil{$^2$\small\it Univ. Grenoble Alpes, CNRS, IPAG, Grenoble, France}
\affil{$^3$\small\it Univ. Grenoble Alpes, CNRS, CNES, IPAG, Grenoble, France}
\affil{$^4$\small\it Jet Propulsion Laboratory, California Institute of Technology, Pasadena, CA, USA}
\affil{$^5$\small\it IPAC, California Institute of Technology, Pasadena, CA, USA}
\affil{$^6$\small\it Univ. of Cologne, Cologne, Germany}
\affil{$^7$\small\it Univ. Paris-Saclay, CNRS, Institut d’astrophysique spatiale, Orsay, France}
\affil{$^8$\small\it Institut d’Optique and Hubert Curien Lab, Univ. de Lyon, Saint-Etienne, France}
\affil{$^9$\small\it Sorbonne Université, CNRS, Institut d’Astrophysique de Paris, Paris, France}
\affil{$^{10}$\small\it Pyxalis, 170 Rue de Chatagnon, 38430 Moirans, France}
\affil{$^{11}$\small\it NASA Goddard Space Flight Center, Greenbelt, MD, USA}
\affil{$^{12}$\small\it Obs. Torino/INAF, Pino Torinese, Italy}



\begin{abstract}
  This study presents a comprehensive system analysis for an instrument onboard the Habitable Worlds Observatory (HWO), designed for high-precision, high-accuracy differential astrometry, with the primary scientific goal to determine the mass of Earth-like planets around the nearest Sun-like stars. The analysis integrates the definition of the mission profile, the instrumental concept architecture, and an error budget that breaks down the key contributors to the sub-µas precision required for a single measurement. A portion of this budget addresses photo-center estimation for both the target and calibration stars used in differential astrometry. Other major contributors are related to instrumental control of systematics in the reconstruction of differential angle measurements from pixel data (focal plane calibration) to on sky line of sight (telescope distortion calibration). End-of-mission astrometry requires multiple observations (typically 100) of the same target distributed over the mission lifetime. We assess the mission profile to estimate the fraction of survey time required for astrometric survey to achieve the science objective. The proposed architecture of the instrument concept is derived from error budget and mission constraints based on a large visible detector array composed of an assembly of multiple CMOS sensor chips resulting in an overall gigapixel focal plane. We evaluate the Technology Readiness Level (TRL) and propose a way forward reaching TRL 5 level for key technologies by the Mission Consolidation Review in 2029.
  \\
  \\
\end{abstract}

\vspace{2cm}

\section{Introduction}

Astrometry plays a key role in the characterization of potentially habitable planets of Earth-like size orbiting around nearby solar type stars. It is a complementary tool to the coronagraphy instrument as well as to radial velocity measurements, for performing astrophysical characterization in the next generation of space missions. While radial velocity is highly effective, it is an order of magnitude more sensitive to stellar activity than astrometry. Astrometry, therefore, plays a crucial role in determining the true mass and three-dimensional orbital parameters of telluric exoplanets around the nearest Sun-like stars. True mass is a key prerequisite for interpreting molecular detections in exoplanets atmospheres and is required to retrieve in an unambiguous manner the composition of the atmosphere (biosignatures) in the spectrum of the reflected light from the candidates’ habitable worlds. The instrument proposed is based on relative astrometric measurements, where the different targets with Earth-like candidates are observed repeatedly over the duration of the full missions. The data collected are composed of relative angular measurements between the target pointed at the center of the instrument and the reference calibration stars within the field of view.
Relative astrometric signal of a star with perturbation from an orbiting exoplanet (in µas) is given by \citep{Malbet2018Astrometry}:
\begin{equation}
S_{\mathrm{astrom}} = 3
\times \left( \frac{M_p}{M_\oplus} \right)
\left( \frac{a_p}{1~\mathrm{AU}} \right)
\left( \frac{M_\star}{M_\odot} \right)^{-1}
\left( \frac{D}{1~\mathrm{pc}} \right)^{-1}
\end{equation}
For the definition of the instrument, we will select the sizing case of a 1 Earth mass planet $M_p$ orbiting a $M_\odot$ star on 1 Astronomical Unit radius orbit ($a_p$) located at D=10 pc from the observer. In that case the astrometric signal is 0.3 µas.
The input data to astrometric signal retrieval is the variation of relative position of the target (star + exoplanet) photocenter in the relative reference frame defined by the calibration stars. The concept of operation relies on multiple sequence of observation spanning several orbiting periods of exoplanets. Each single observation consists of one pointing of the telescope to a target and staring for a duration of $< 1~\mathrm{hour}$. Some additional calibrations are taken for each sequence to complete the data set.
The input data to astrometric signal retrieval is the variation of relative position of the target (star + exoplanet) photocenter in the relative reference frame defined by the calibration stars. The concept of operation relies on multiple sequence of observation spanning several orbiting periods of exoplanets. Each single observation consists of one pointing of the telescope to a target and staring for a duration of $< 1$ hour. Some additional calibrations are taken for each sequence to complete the data set.
The typical number of revisits assumed for the mission definition is $N_{\mathrm{obs}}$=100 revisits per target. The distribution of the target sequence of observation will depend on the pointing capability of the observatory accounting for limitation of impacts of Solar Aspect Angle variation on thermal stability and straylight.
The end of mission Signal to Noise ratio of the astrometric signal: \begin{equation}
\mathrm{SNR}_{\mathrm{astrom}}
= \sqrt{N_{\mathrm{obs}}} \times 
\frac{S_{\mathrm{astrom}}}{\sigma_{\mathrm{1\,meas}}}
\end{equation}
is set to a minimum of 6 \citep{Malbet2025HWO} leading to a single measurement error requirement $\sigma_{\mathrm{1\,meas}} = 0.50~\mu\mathrm{as}$. This is the error for which we will size the astrometric instrument.

\section{Budget error}

\subsection{Budget allocation}
Starting from the single differential error budget of 0,5 µas the allocation is broken down to the following main contributors:
\begin{enumerate}[label=\alph*)]
\item Error of photocenter determination of target and calibration stars
\item Error of pixel position to on-sky line of sight calibration
\item Astronomical errors
\end{enumerate}
The following budget allocation is proposed (Fig. 1) and justified in following sections 2.3 and 2.4 by bottom-up verification of performance (astronomical errors are not discussed in this paper):
\begin{figure}[ht]
\includegraphics[width=\columnwidth]{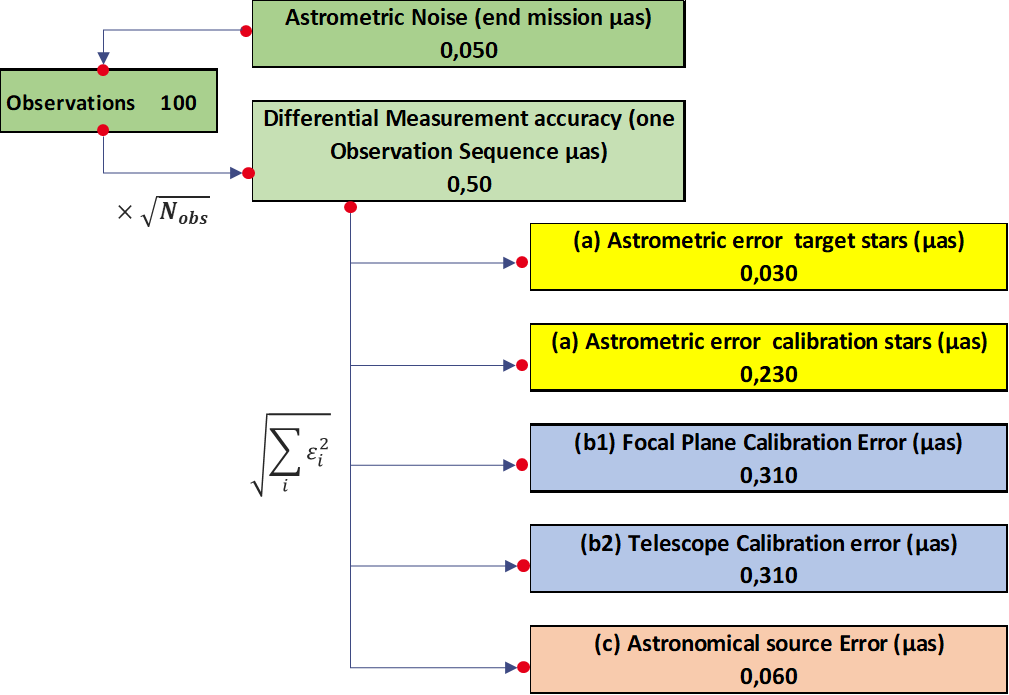}
\caption{\small Main budget allocation for high-accuracy high-precision Astrometry.
\label{author_fig1_label}
}
\end{figure}

All of these contributors are at the sub-micro arcsec levels making the end of mission performance mainly a calibration experiment to control the systematics below the statistical errors floor.
\subsection{Assumption for design analysis}
The telescope will be assumed to be the one defined in Engineering Architecture Concepts EAC3 configuration with a diameter of 7,2 m and a pupil composed of 19 hexagons with a collecting area of 32,8 m² (effective diameter 6,5 m).
The optical performance of the system will be characterized at first order by the FWHM of the PSF $ \mathrm{FWHM}_{\mathrm{PSF}} = 1.03 \times \lambda / D_{\mathrm{tel}} = 20~\mathrm{mas} $ in the visible waveband (mean wavelength = 680 nm) under assumption that the system is close to diffraction limit as is expected for a high-performance coronagraph mission.
The design analysis is based on \citet{mamajek2024nasaexoplanetexplorationprogram} defining a list of preliminary representative sample of the targets that need to be addressed by the HWO mission. This list gives a sizing scenario for the observation sequence.
For the photometry of target stars, based on the sample from the preliminary target list, we assume magnitude ranging from 3 to 7 with a  reference Spectral Energy Distribution corresponding to a G5V star $T_{\mathrm{eff}} = 5660~\mathrm{K}$ and $R_\star = 0.997 \times R_\odot$.
\subsection{Photon noise on photocenter determination}
The fundamental limit for astrometry accuracy is related to centroid measurement of image of stars in the pixelized focal plane. Best method to determine a stable photocenter associated to the image of the stars is still under study but for all methods, the ultimate accuracy is inversely proportional to the SNR of star cumulated flux limited by photon noise \citep{Guyon2012ApJS}.
For the target star, the contribution to the final astrometric error is given by: \begin{equation}
\sigma_{\mathrm{ast,target}} =
\frac{\mathrm{FWHM}_{\mathrm{PSF}}}{2 \times \mathrm{SNR}_{\mathrm{target}}}
\end{equation}
with
\begin{equation}
\mathrm{SNR}_{\mathrm{target}} =
\frac{N_{\mathrm{target\,photo\text{-}e^-}}}
{\sqrt{N_{\mathrm{target\,photo\text{-}e^-}}}}
\end{equation}
For the calibration stars, assuming that the relative astrometric measurement is based on $N_{\mathrm{cal_star}}$ pairs of angle estimations, the contribution to the final astrometric error is:
\begin{equation}
\sigma_{\mathrm{ast,cal}} =
\frac{1}{\sqrt{N_{\mathrm{cal\,star}}}} \times
\frac{\mathrm{FWHM}_{\mathrm{PSF}}}{2 \times \mathrm{SNR}_{\mathrm{cal\,star}}}
\end{equation}
with:
\begin{equation}
\mathrm{SNR}_{\mathrm{target}} =
\frac{N_{\mathrm{calibration\,photo\text{-}e^-}}}
{\sqrt{N_{\mathrm{calibration\,photo\text{-}e^-}}}}
\end{equation}
In the case of HWO mission, with a large primary mirror (small PSF) and large collecting area it is possible to reach the required allocation for both target and calibration stars in an observation sequence of less than 2500 s (see Fig. 2).
\begin{figure}[ht]
    \centering
    \includegraphics[width=\columnwidth]{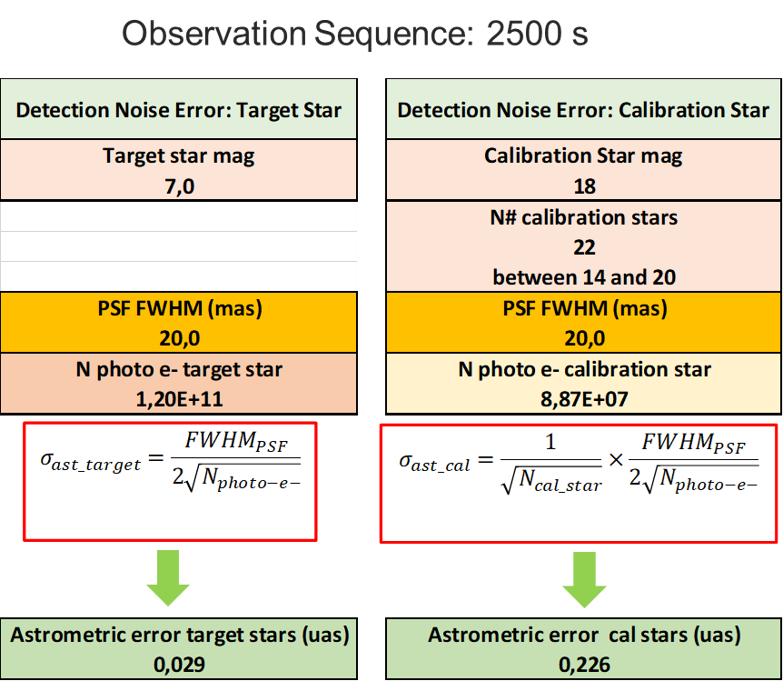}
    \caption{\small Estimated photon-noise astrometric error budget for the faintest target (mag 7) and minimum number of calibration stars (22 stars with an average magnitude of 18).}
    \label{author_fig2_label}
\end{figure}

This contribution corresponds to the performance linked to the primary parameters of the mission, mainly the telescope collection area, the sharpness of the PSF, the cumulative integration time on the target and represents ultimate performance accessible to the mission. All other contributors are linked to secondary order parameters, meaning post-calibration allowing control of bias in the measurements as described in the following sections.
\subsection{Line of sight Calibration}
The centroid measurement of target and calibration stars is performed in the instrument focal plane pixelized space. This measurement must be converted to relative astrometric measurement meaning to angular difference between directions expressed at entrance of the telescope (as illustrated in Fig. 3).
This requires two levels of calibration within the instrument and one level of astronomical calibration from observer to celestial sphere (see section 2.5) that are identified as contributors b) and c) in the budget (see Fig. 1).
\begin{figure}[ht]
\includegraphics[width=\columnwidth]{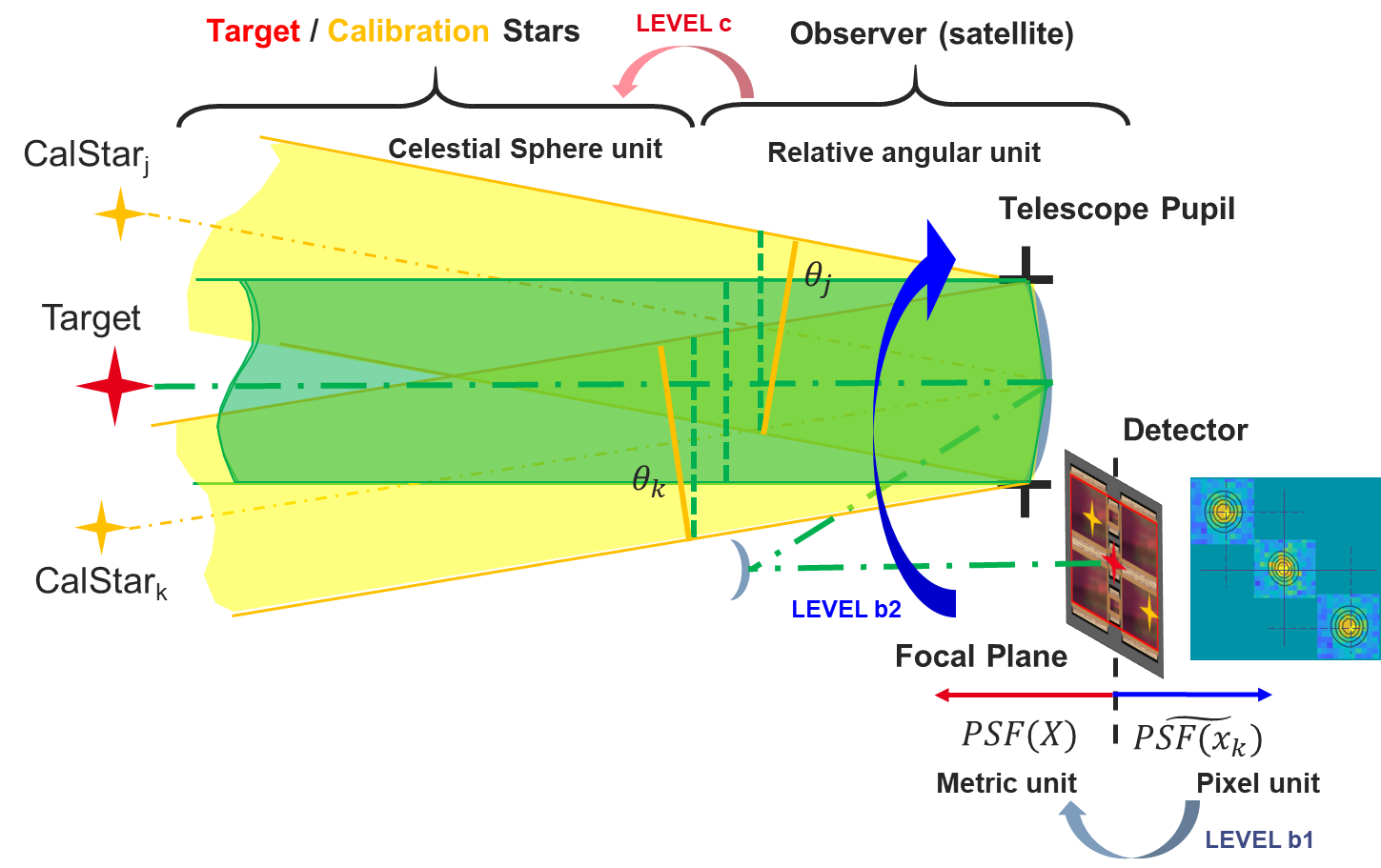}
\caption{\small Illustration of calibration levels from measured centroid in pixel space to angular relative astrometry.
\label{author_fig3_label}
}
\end{figure}

First level (level b1) is the focal plane pixel grid calibration: relating a measurement in the pixel grid to an optical position in the telescope image focal plane (before pixelization).
Because of variation of pixel response to flux illumination over CMOS array (inter-pixel variation) due to manufacturing and physical properties dispersion, sub-µas astrometry requires calibration of the full-array to correct for at least the flat-field response, and the effective geometric pixel centroid shift from a regular grid.
It has been reviewed in \citet{Zhai2011PRSA} that this calibration can be obtained by measuring the signal of pixels exposed to sinusoidal illumination pattern generated by fringes from interference of two distant laser metrology beams (Young’s fringes pattern). The signal retrieval can be enhanced by scanning the fringes over the focal plane introducing offset between the two interfering lasers frequency.
Laboratory calibration to the level of $3 \times 10^{-5}$ pixels to $6 \times 10^{-5}$ pixels have been achieved by teams in JPL \citep{Shao2023PASP} and Grenoble \citep{Crouzier2016AA}.
At first order the calibration error is estimated based on the SNR of cumulated flux integrated in the fringes during the calibration sequence by \citep{Zhai2011PRSA}:
\begin{equation}
\sigma_{\mathrm{pixel,cal}} =
\frac{1.5 \times \mathrm{Pixel}_{\mathrm{on\,sky}}}
{\mathrm{SNR}_{\mathrm{fringe}}}
\end{equation}
with: 
\begin{equation}
\mathrm{SNR}_{\mathrm{fringe}} =
\frac{N_{\mathrm{fringe\,photo\text{-}e^-}}}
{\sqrt{N_{\mathrm{fringe\,photo\text{-}e^-}}}}
\end{equation}
For this step an internal calibration unit is designed (see section 4.1) and is part of the on-board instrument. The calibration budget is expected to be reached in a 1400-s calibration sequence associated to each observation sequence (see Fig. 4).
\begin{figure}[ht]
\includegraphics[width=\columnwidth]{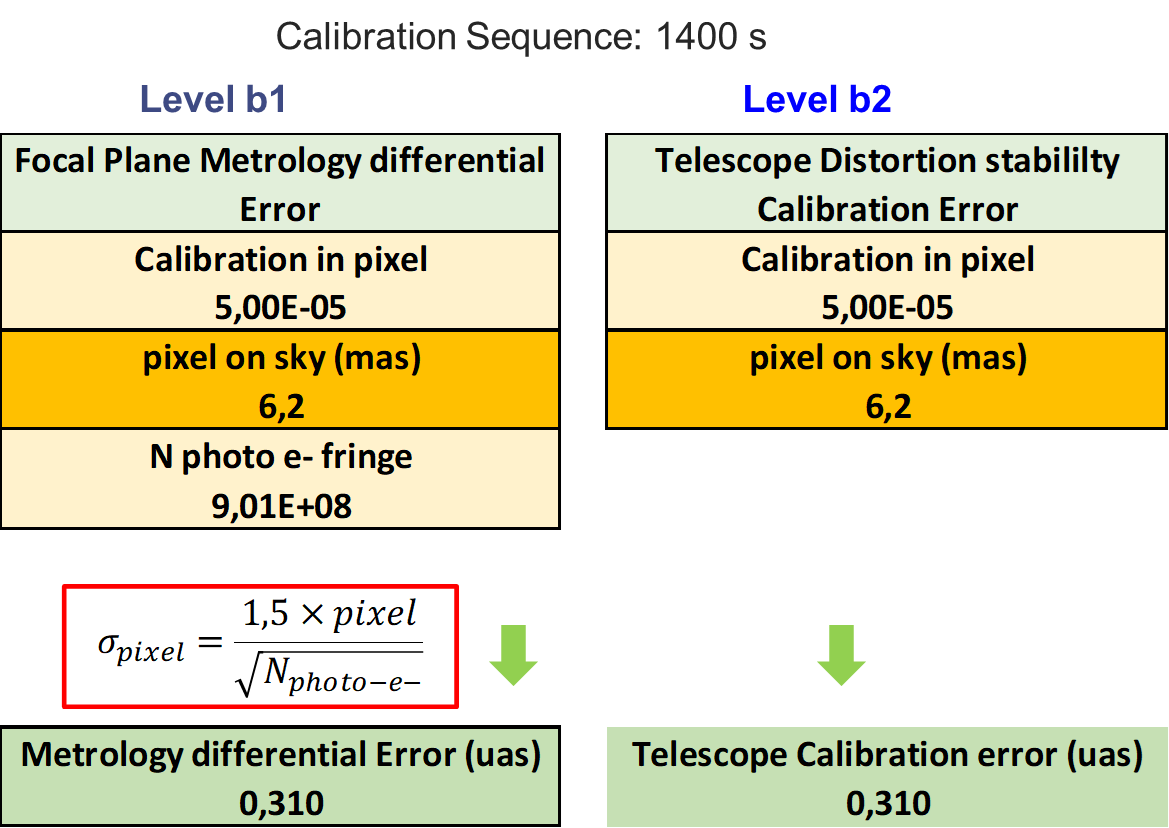}
\caption{\small Instrumental Calibration budget.
\label{author_fig4_label}
}
\end{figure}
Second instrument calibration (level b2) is the telescope optical grid calibration: relating a measurement in the telescope image focal plane to an angular direction at entrance of telescope.
The centroid of the optical PSF is sensitive to small variations of the Wave Front Error along the telescope optical path. Though Astrometric instrument will benefit from the very high thermal stability of the HWO payload on short time-scales (typically that of an observation sequence), the longer timescales variations need to be calibrated out.
In particular, the same calibration stars used to define the relative reference frame for each target will be observed at different field positions during the survey (due to limitations on pointing capability of the spacecraft) leading to variable distortion as the optical footprint of each field will sample different area of the optical elements. In addition, the long timescale thermal elastic deformation might induce optic WFE deformations beyond what is acceptable for the current budget allocation.
Some previous analysis (\citet{Shao2023PASP}, \citet{lizzana2024experimentaltestscalibrationhigh}) indicate that the optical distortion function can be modeled to the required accuracy using a two-dimensional polynomial model calibrated by using reference stars in the FoV.
This point remains a challenge to be demonstrated on more realistic optical systems and we will assume at that stage that the final calibration will rely on several steps of data processing starting from ground characterization to in-flight calibration. The full optical chain needs to be calibrated on dense star field with construction of a model linking thermal behavior of the telescope to optical quality. The spectral wavelength behavior of the PSF will need to be taken into account as the observations are broadband. In addition, during each sequence of observation, the optical model needs to be updated using the in-field reference stars with possibly some scanning (dither or roll) over the focal plane.
Some additional calibration approach might be required to reach the high performance for relative astrometry. In particular the use of diffractive pupil pattern is currently investigated (\citet{Guyon2012ApJS}, \citet{Bendek2023SPIE}). Several options are being studied, using the diffractive pattern from the M1 mirror segmentation or specific dots pattern on the M1 mirror to calibrate the full optical chain. Or using a diffractive pupil pattern in the instrument, calibrating only the instrument part and relying on stability for the telescope part. In any case, the astrometric instrument will benefit from minimizing the optical elements, and in particular the refractive elements.
\subsection{Astronomical errors}
The last level of calibration (level c) requires to correctly account for astronomical contributions in particular stars proper motions, variable parallax at different epoch of observations and relativistic correction of line of sight between the observer (satellite defined by attitude and orbit) and the celestial objects. Though astrometry is 10 times less sensitive to starspots than radial velocity method stellar activity impact on astrometric measurement needs to be taken into the error budget.
As the instrument is using broad band visible imaging, the spectral energy distribution of the stars needs to be characterized to a level compatible with the astrometric performance.
\section{Instrument Design}
The Astrometric instrument design is driven by the following main needs:
\begin{enumerate}[label=\alph*)]
\item High spatial resolution: sampling of Optical PSF to reach statistical floor on photon noise;
\item Large Field of view to maximize the number of reference calibration stars for relative astrometry;
\item Large Dynamics, Fast Read out to accommodate large range of observed object magnitude.
\end{enumerate}
\subsection{High spatial resolution}
The waveband selected for the instrument is ranging from 420 nm to 940 nm.  The instrument is sized to sample the FWHM of the smallest PSF (at minimum wavelength, see Fig. 3) with at least 2 pixels.
\begin{figure}[ht]
\includegraphics[width=\columnwidth]{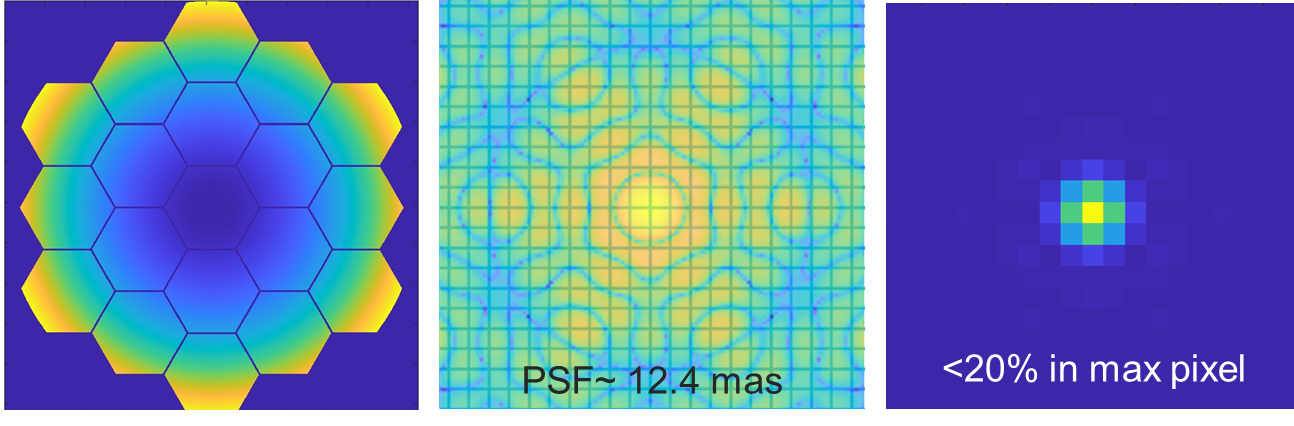}
\caption{\small Engineering Architecture Concepts EAC3 Primary mirror telescope and associated PSF at 420 nm.
\label{author_fig5_label}
}
\end{figure}

The FWHM at 420 nm is 12.4 mas, so pixel on sky shall have a size of maximum 6.2 mas. It is a fundamental pre-requisit for extreme astrometry to have a correctly sampled PSF to prevent pixel phase error that would prevent reaching ultimate performance for centroid estimate (see \citet{Anderson2000PASP}).
\subsection{Large Field of view}
The size of the field of view is defined by the number of calibration stars that need to be used for relative astrometric measurement.
In the current allocation, the photon noise level for centroid error associated to calibration stars can be reached for various couple of number of calibration star available in field $N_{\mathrm{cal_star}}$ and average star magnitudes for a given integration time during a sequence of observation. For the reference case of integration time = 2500-s, Fig. 6 provides the minimum required number of stars for a given average magnitudes within the field of view.
\begin{figure}[ht]
\includegraphics[width=\columnwidth]{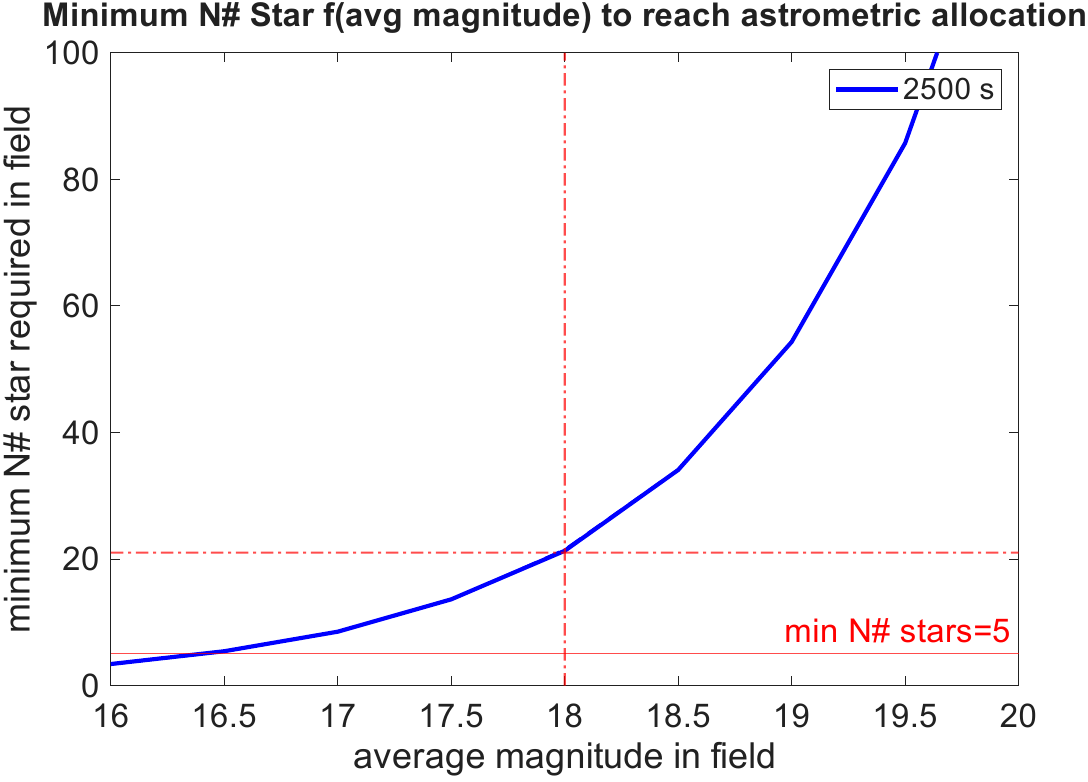}
\caption{\small Minimum number of calibration stars required in a field function of average magnitude of stars in the field to reach required allocation on error from Calibration Star.
\label{author_fig6_label}
}
\end{figure}
We have analyzed the 164 targets provided in \citealt{mamajek2024nasaexoplanetexplorationprogram}. For each of them we access the GAIA DR3 database \citep{Lindegren2021AA} to review the stars present in the close field of the target. We reject the objects for which the re-normalised unit weight error (ruwe) parameter is $> 1.4$ as an indicator for quality assessment of GAIA DR3 data. We then compute, as a function of increased size of the field, the average star magnitude and the number of stars available, we set a minimum number of 5 stars to allow relative astrometry. We can then estimate the fraction of HWO potential targets for which the relative astrometric performance allowing detection of Earth-like planet is reached.
The assessment is performed for two typical fields, one of $2 \times 3~\mathrm{arcmin}^2$ corresponding to the field allocation of the HRI instrument and one $> 3 \times 4~\mathrm{arcmin}^2$ corresponding to a dedicated astrometric instrument located at one of the FGS fields of Engineering Architecture Concepts EAC3 (maximum field there would be $3 \times 6~\mathrm{arcmin}^2$).
While a 6 $\mathrm{arcmin}^2$ allow to reach performance on $< 50\%$ of the targets, a 12 to 18 $\mathrm{arcmin}^2$ allow reaching performance on $65\%$ to $80\%$ of the HWO preliminary target (see Fig. 7) using calibration stars down to magnitude 20.
\begin{figure}[ht]
\includegraphics[width=\columnwidth]{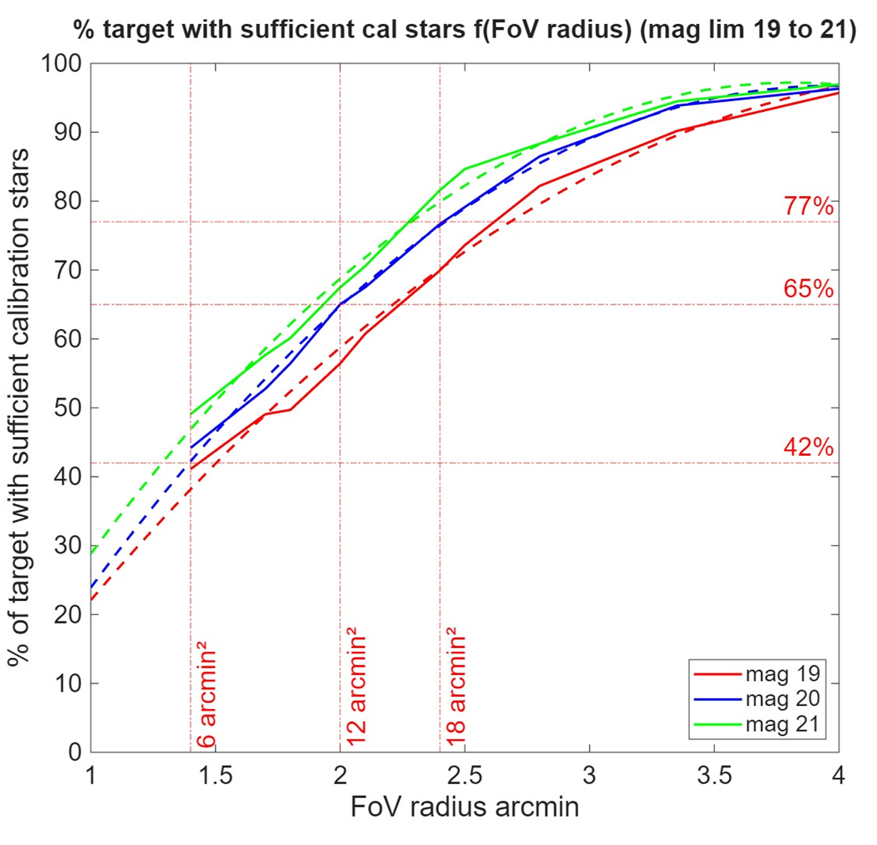}
\caption{\small Fraction of targets for which the budget performance is reached as a function of the astrometric instrument FoV.
\label{author_fig7_label}
}
\end{figure}

Setting the field of view of the instrument to $3 \times 4~\mathrm{arcmin}^2$ with a pixel size on sky to 6.2 mas leads to a focal plane size of $29000 \times 39000$ pixels so a gigapixel focal plane array.
The main improvement in completeness of target coverage is gained by increasing the Field of View, while increasing the observation sequence brings little improvement as the number of stars available in small fields are very limited. For a 6 $\mathrm{arcmin}^2$ FoV, the completeness tends to saturate at around $50\%$ of target list even when increasing integration time.
\subsection{Large dynamics and Read out}
The design of the instrument is relying on CMOS detectors in the GIGAPYX CMOS family developed by Pyxalis. This family of large CMOS sensor have been developed from the inception as an evolutive format manufactured using stitching technology. This ensures that different format shares similar pixel and readout blocks and also similar performance.
One of the instrument drivers is to be able to detect an extremely large dynamics of star signal, with target stars ranging from magnitude 3 to 7, and with calibration stars of magnitude as low as magnitude 20.
From those constraints, the concept of instrument (Fig. 8) has been designed based on a visible focal plane array composed of an assembly of four 220 Mega Pixels (MP) 4.4 µm pixel CMOS sensor chip leading to a gigapixel scale focal plan with a central small field (smaller CMOS detector $< 14$ Mega Pixel MP).
\begin{figure}[ht]
\includegraphics[width=\columnwidth]{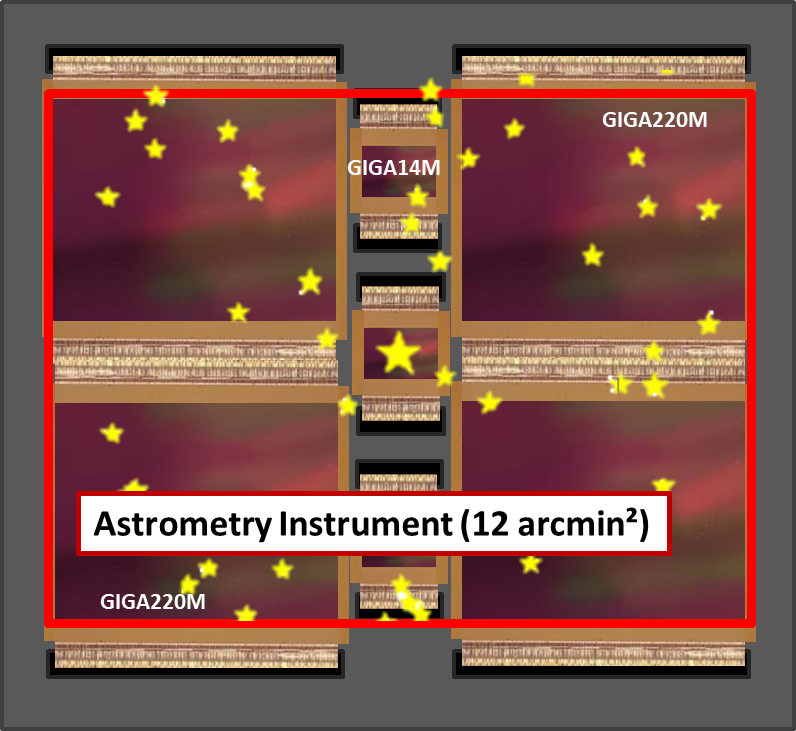}
\caption{\small Illustration of a $3 \times 4~\mathrm{arcmin}^2$ instrument composed of 4×220 MegaPixel and 3×14 MegaPixel CMOS (for redundancy) with target and calibration stars corresponding to one exemple of HWO preliminary target HD146233 (18 Scorpi).
\label{author_fig8_label}
}
\end{figure}
The central detector is dedicated to centroid measurements of the star+exoplanet bright target toward which the telescope is pointing while the wider field CMOS detectors array around are dedicated to the measurements on the multiple faint calibration stars. This architecture allows covering the large required FoV and the large dynamics of targets. The 4.4 µm pixels have a full well capacity of 50 000 electrons and the detector Quantum Efficiency is illustrated on Fig. 9.
\begin{figure}[ht]
\includegraphics[width=\columnwidth]{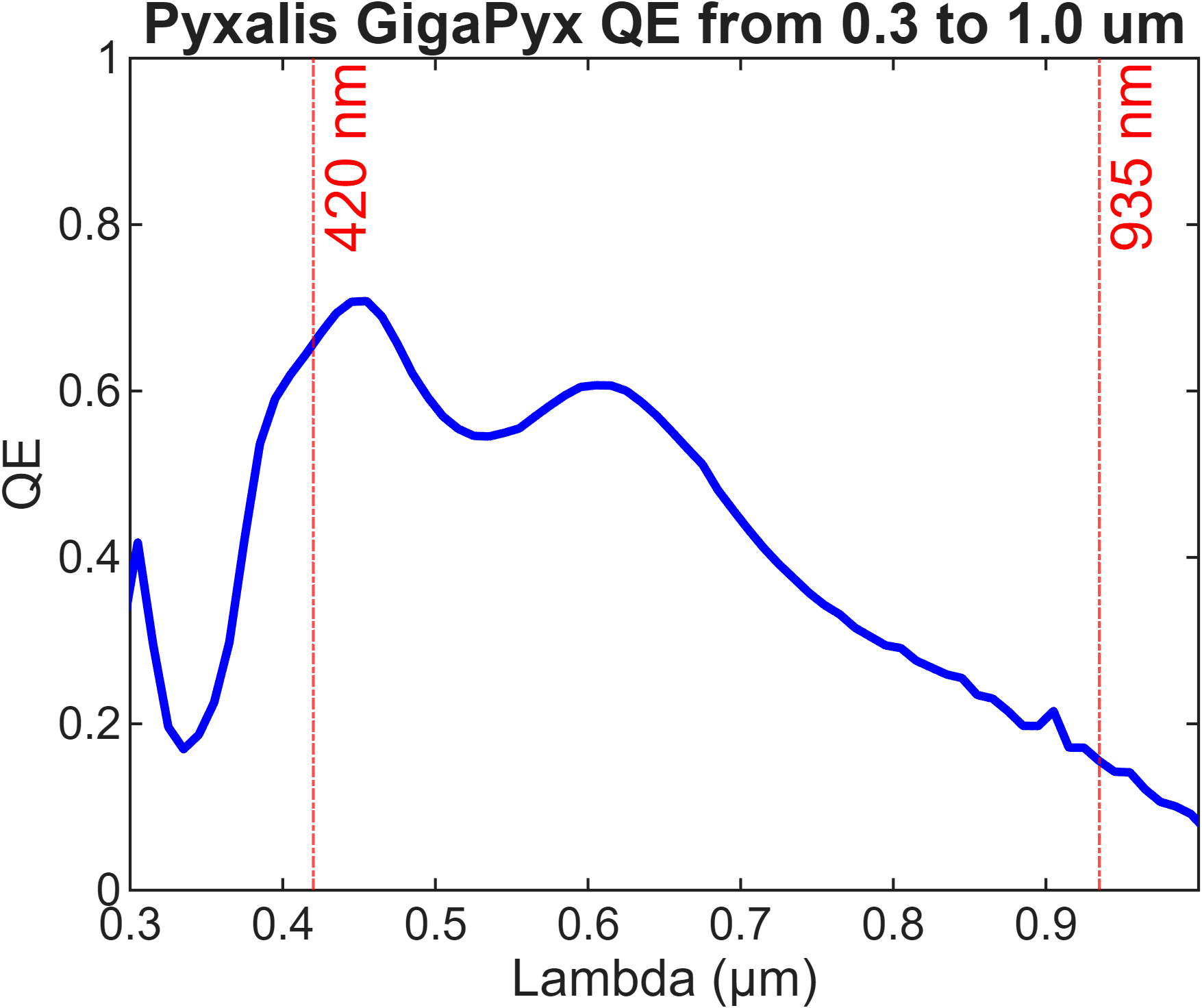}
\caption{\small GIGAPYX CMOS Quantum Efficiency.
\label{author_fig9_label}
}
\end{figure}
The bright target will saturate the Full Well in less than a few ten’s of µs while the calibration stars can be integrated for hundreds of ms before saturating. Pyxalis has designed two possibilities of readout modes. The standard readout mode is used for the large CMOS (fainter calibration targets) with high photo-electron collecting efficiency (~$< 90\%$). It allows reading the 220 MP in full frame at around 3 to 10 frames per second.
The fast readout mode will be used on the small 14 MP detector dedicated to the bright target detection. It allows reading a subarray centered on the target star in window mode with integration time as small as a few microseconds and window rate of up to 5000 frame per seconds preventing saturation in very bright conditions, at the cost of having less efficient photo-electron collection on sky (which is not problematic in the case of the bright targets).
This configuration of focal plane designed to handle large signal dynamics is also suitable to detect diffractive patterns in the full field for optical calibration as described in \citet{Bendek2023SPIE}.

\section{Baseline implementation}
\subsection{Instrument Architecture}
The baseline architecture of the instrument (Fig. 10) is based on the GIGAPYX detector family with a visible focal plane composed of $4 \times 220$ MP CMOS “calibration stars” arranged in the outer field and $3 \times 14$ MP CMOS “target” in the central field. Though only 1 small “target” detector is required, a set of 3 small detectors will be arranged in the central field in order to prevent loss of instrument due to single point of failure.
The mechanical design is based on an on-going evolution of the GIGAPYX CMOS toward an Astro packaging version designed to ensure thermal mechanical and electrical interfaces compatible with space qualification program. In particular the routing of the numerous electrical signals toward the electronics control unit is under definition.
\begin{figure}[ht]
\includegraphics[width=\columnwidth]{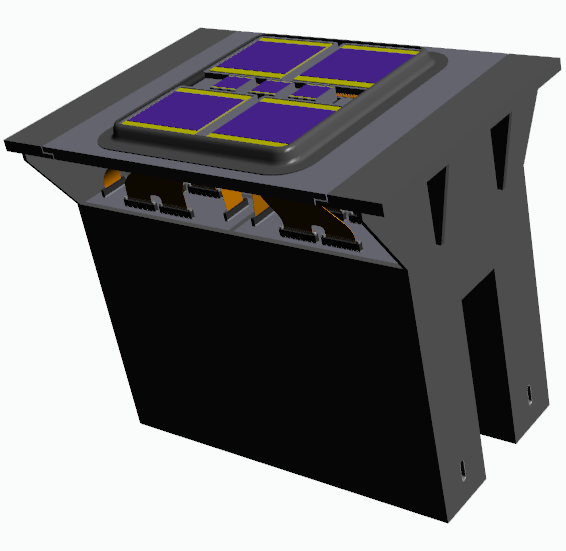}
\caption{\small Astrometric instrument focal plane and electronics unit.
\label{author_fig10_label}
}
\end{figure}
In order to ensure correct PSF sampling and size of field of view with 4.4 µm pixels, the instrument focal plane is assumed to be located at one of the EAC3 telescope focal plane for the FGS instrument with a focal length of ~150 m and a f-number of 20.3. 
The metric size of the focal plane instrument in this configuration is estimated to a volume of $200 \times 300 \times 350 \mathrm{mm}^3$ for a mass of 20 kg. The dissipated power at the focal plane level is estimated to be 15 W. The current baseline operation temperature is set between 270 and 290 K (detectors are qualified down to 233 K).
A Focal Plane Calibration unit is designed to project Young’s fringes pattern on the focal plane from pairs of monomode fibers connected to an HeNe Laser. It is based on development of a laboratory experiment at IPAG Grenoble to validate the focal plane metrology. The Calibration unit concept is based on integrated optical components largely developed for optical communication (including space communication) and rely on the following on-board elements (Fig. 9):
\begin{enumerate}[label=\alph*)]
\item HeNe Laser source light is split into two fibers;
\item Modulation applied by two lithium niobate phase modulators (periodic phase shift between the interfering beams);
\item A multi-fiber configuration along two axis (Horizontal and Vertical bases) allowing via a switch to project combination of two fibers fringes on the Focal Plane.
\end{enumerate}
The pixel center offsets are obtained by comparing the phase of the sinewaves observed on individual pixels with the phase of the global system of fringes.
Laser and opto-electronics functions can be located on the spacecraft warm electronics area while the fibers are routed toward the last mirror in front of the instrument focal plane (Fig. 11).
\begin{figure}[ht]
\includegraphics[width=\columnwidth]{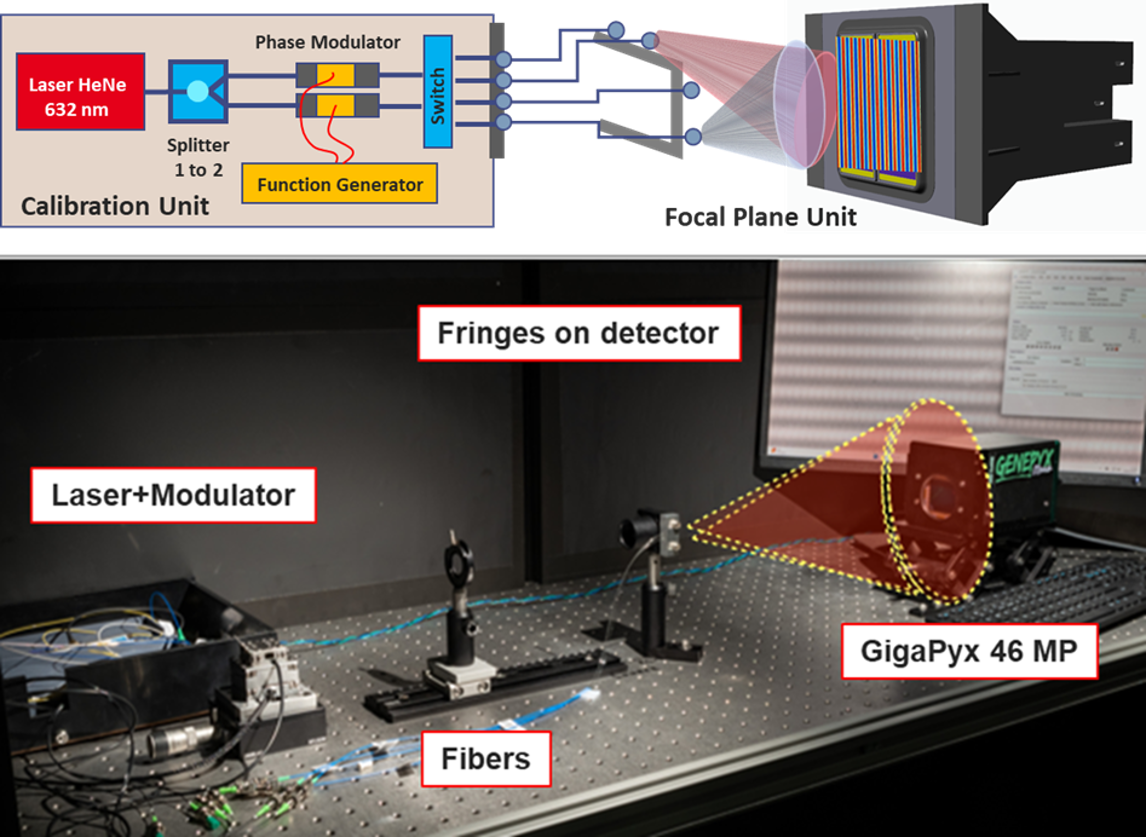}
\caption{\small Pixel Fringe Calibration unit principle (top) and laboratory test at IPAG (bottom).
\label{author_fig11_label}
}
\end{figure}

Based on laboratory hardware the calibration unit is allocated 8 kg and a power dissipation of 20 W.
\subsection{Laboratory test bed and TRL increase}
A 46 MP GIGAPYX CMOS detector is currently being tested at IPAG Grenoble for assessment of performance (read out noise, dark current, PRNU) and calibration of pixel metrology \citep{Lizzana2025HWO}. This device has already been submitted to radiation testing in order to validate the compatibility of the design with space environment \citep{michelot2025gigapyxsensorperformancespace}. A joint development between IPAG, CEA and Pyxalis funded by CNES and with additional funding proposal on-going aims at manufacturing a 220 MP CMOS with a first batch expected beginning of 2026.
One important part of the current development is to integrate the CMOS sensor into a packaging (ASTRO package see Fig. 12) which is suitable for larger focal plane assembly of multiple detectors adapted for space environment constraints (vibration loads, thermal control, mechanical alignment) and interface toward dedicated space readout electronics. Pyxalis will be supported at that stage by CEA expertise on detector integration in large focal plane array for space instrument (Euclid VIS Focal Plane Assembly).
\begin{figure}[ht]
    \centering
    \includegraphics[width=\columnwidth]{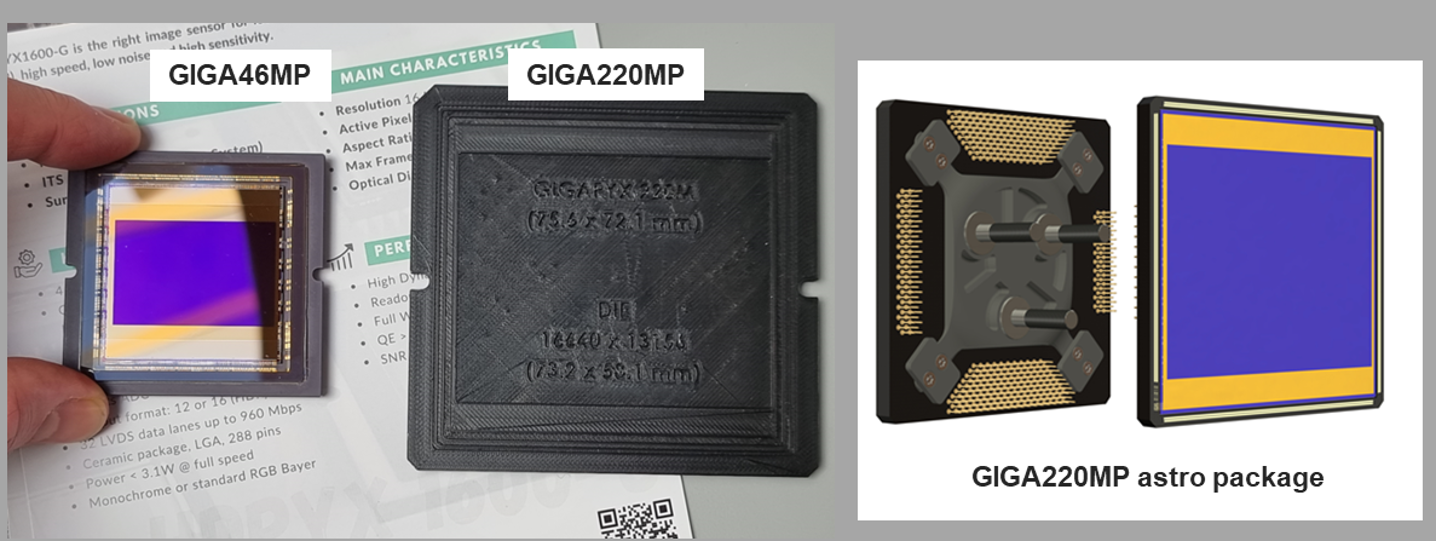}
    \caption{\small Increase of the Technology Readiness Level (TRL) of the GIGAPYX family from 46~MP toward GIGAPYX~220~MP with the Astro package.}
    \label{author_fig12_label}
\end{figure}

The on-going work covers the realization of a full scale 12 arcmin² focal plane with $2 \times 220$ MP CMOS manufactured by Pyxalis, and the associated readout and control electronics developed by CEA. Pending on funding availability, the work plan will cover integration of a focal plane with 2 engineering models of CMOS, and two mechanical models and testing of the fringe projection for calibration of large focal plane array with a goal to reach a TRL of 5 on components and TRL of 3 on instrument in 2029. The main activities for TRL increase will cover the identified critical items for the instrument which are: large CMOS detectors with space qualified packaging, pixel grid calibration on-board unit validation on large field, on-board processing of large data volume produced by gigapixel focal plane.
\section{Conclusion}
A proposal for an astrometry instrument on-board HWO is derived from requirements to detect relative astrometric signal from an Earth-like planet around a sun-like star at 10 parsec leading to sub-µas error allocation to the instrument.
The sequence of observation proposed rely on 2500 seconds of single observation per target with an additional 1400 seconds for fringe calibration of the pixel grid. This sequence is repeated 100 times along the mission lifetime for each target. The observation of 60 targets would require ~$15\%$ of a 5-years survey lifetime and observation of the 164 preliminary targets would require $< 10\%$ of the 25-years HWO non-serviceable lifetime. 
The proposed instrument is based on Gigapixel CMOS focal plane array covering a 12 $\mathrm{arcmin}^2$ field of view with two types of detectors: $4 \times 220$ MP arrays with normal readout (3 to 10 fps) to detect the calibration stars signal, and $3 \times 14$ MP central array with fast window readout (up to 5000 fps) to detect the bright target.
The focal plane array is associated to a focal plane calibration unit projecting Young’s fringes on the detectors to calibrate the pixel grid and flat field.
The critical issue of optical distortion calibration is under study to propose a suitable, multi-stage calibration approach.
A 46 MP CMOS is under characterization at IPAG and a plan to develop the 220 MP CMOS in an Astro packaging suitable for space mission is on-going with a goal to have an integrated focal plane prototype including detectors and readout electronics in 2029 with validation of the calibration method.

{\bf Acknowledgements.} We acknowledge the HWO community and the many contributors to concept studies that motivated this work.
JA thanks collaborators and institutions listed in the author affiliations for support and discussions.
With regard to the funding of our research, we would like to acknowledge the support of the LabEx FOCUS ANR-HWO Astrometry for Exoplanets and Dark Matter 7 11-LABX-0013 and the CNES agency. ML would like to acknowledge the support of her PhD grant from CNES and Pyxalis.
This research has made use of NASA Astrophysics Data System Bibliographic Services and of data obtained from or tools provided by the portal exoplanet.eu of The Extrasolar Planets Encyclopaedia.

\bibliography{author.bib}

\end{document}